\documentclass[12pt,twoside]{article}   
\usepackage{epsfig,times}  % 17 dec. 2001: pbs avec times
\usepackage{color}

\begin{document}
\thispagestyle{empty} 

%\markboth{{ \sl \hfill Chapter 1 \hfill \ }}
%         {{ \sl \hfill Introduction \hfill \ }}

%%%%%%%% Pour changer les valeurs par defaut pour taille figure,
%%%%%%%% sinon au-dela d'une hauteur de 134 mm = 70% on est rejete a la fin
 \renewcommand{\topfraction}{.99}      
 \renewcommand{\bottomfraction}{.99} 
 \renewcommand{\textfraction}{.0}

%%%%% Definitions

\newcommand{\nc}{\newcommand}

\nc{\qI}[1]{\section{{#1}}}
\nc{\qA}[1]{\subsection{{#1}}}
\nc{\qun}[1]{\subsubsection{{#1}}}
\nc{\qa}[1]{\paragraph{{#1}}}

            % Enumerations
\def\qbu{\hfill \par \hskip 6mm $ \bullet $ \hskip 2mm}
\def\qee#1{\hfill \par \hskip 6mm #1 \hskip 2 mm}

\nc{\qfoot}[1]{\footnote{{#1}}}
\def\qL{\hfill \break}
\def\qpar{\vskip 2mm plus 0.2mm minus 0.2mm}
\def\tvi{\vrule height 12pt depth 5pt width 0pt}
\def\qtvi{\vrule height 2pt depth 5pt width 0pt}
\def\qth{\vrule height 15pt depth 0pt width 0pt}
\def\qtb{\vrule height 0pt depth 5pt width 0pt}

\def\qparr{ \vskip 1.0mm plus 0.2mm minus 0.2mm \hangindent=10mm
\hangafter=1}

                % Decale UN paragraphe
                % Attention! La double accolade est vitale, sinon tout le
                % est decale (cf TEX p.199)
                % On peut aller a la ligne avec \qL=\hfill \break
                % Par contre ne supporte pas les lignes blanches
\def\qdec#1{\par {\leftskip=2cm {#1} \par}}
          % boldface + blue
\def\qbfb#1{\bf\color{\blue}{#1} }

   %% Defs specifiques
\def\qdpt{\partial_t}
\def\qdpx{\partial_x}
\def\qddpt{\partial^{2}_{t^2}}
\def\qddpx{\partial^{2}_{x^2}}
\def\qn#1{\eqno \hbox{(#1)}}
\def\qds{\displaystyle}
\def\qw{\widetilde}
\def\qmax{\mathop{\rm Max}}   % Petit livre Tex (p.167)
\def\qmin{\mathop{\rm Min}}   % Petit livre Tex (p.167)

\def\qs#1{{\bf \color{blue} \LARGE {#1}}\quad }

   %% Pour les legendes des figures
\def\qv{\vskip 0.1mm plus 0.05mm minus 0.05mm}
\def\qhu{\hskip 1mm}
\def\qhv{\hskip 3mm}
\def\qvv{\vskip 0.5mm plus 0.2mm minus 0.2mm}
\def\qhw{\hskip 1.5mm}
\def\qleg#1#2#3{\noindent {\bf \small #1\qhw}{\small #2\qhw}{\it \small #3}\qv }

%%%%% End of definitions

\large

\centerline{\Large Death rate during the exponential growth of {\it E. coli}}
\vskip 8mm
Lo\"{\i}c Brot$ ^1 $, 
Zengru Di$ ^2 $,
Romain Morichon$ ^3 $,  
\qL
Annie Munier-Godebert$ ^4 $,
Bertrand Roehner$ ^5 $, 
Jo\"elle Sobczak-Th\'epot$ ^6 $,
\qL
Ang\'elique Vinit$ ^7 $

\vskip 5mm
\centerline{\it \small Version of 10 December 2021}
\centerline{\it \small Very provisional. Comments are welcome.}
\vskip 3mm

{\small Key-words: Escherichia coli, cell division, mortality
rate, exponential phase, stationary phase} 

\vskip 3mm

{\normalsize
1: Centre de Recherche Saint Antoine (CRSA),
Sorbonne Universit\'e, 75012 Paris, France.\qL
Email: loic.brot@sorbonne-universite.fr \qL
2: School of Systems Science, Beijing Normal University, China.\qL
Email: zdi@bnu.edu.cn\qL
3:  Centre Cytom\'etrie Imagerie Saint-Antoine (CISA).
Sorbonne Universit\'e, 75012 Paris, France\qL
Email: romain.morichon@sorbonne-universite.fr\qL
4:  Centre Cytom\'etrie Imagerie Saint-Antoine (CISA).\qL
Sorbonne Universit\'e, 75012 Paris, France\qL
Email:  annie.munier\_godebert@upmc.fr\qL
5: Institute for Theoretical and High Energy Physics (LPTHE),
Pierre and Marie Curie campus, Sorbonne University,
Paris, France. \qL
Email: roehner@lpthe.jussieu.fr\qL
6: Sorbonne Universit\'e, National Institute for Health and Medical
Research (INSERM).\qL
Email: jsobczakt@gmail.com \qL
7:  Centre Cytom\'etrie Imagerie Saint-Antoine (CISA).\qL
Sorbonne Universit\'e, 75012 Paris, France\qL
Email: angelique.vinit@sorbonne-universite.fr
}
\vfill\eject
%---------------------------------------

{\bf Abstract} \qL

So fast is the growth of a culture of {\it E. coli} that 
it led researchers to overlook a possible death rate.
As a matter of fact,
the experiments done in the first half of the 20th century
were unable to detect any mortality. It is only at the
beginning of the 21th century that positive evidence
emerged which confirmed the existence of a 
sizeable (albeit small) mortality. In the present paper
this mortality was measured at successive 45mn time intervals
from early exponential growth into the stationary phase.
Done with a flow cytometer, the successive  measurements of
the ratio dead/living
also provided an estimate of the standard deviation of the
mortality rate.
In a forthcoming experiment we are planning to focus
more closely on the stationary phase in order to
find out if its mortality rate is lower or
higher than in the exponential phase.

\vfill\eject

%--------------------------

\qI{Introduction}

\qA{Reasons for measuring the mortality rate
of bacteria}

In any population, whether human or non-human, the birth
and death rates are the two most basic variables which define
its evolution. Historically, these were the first variables
which could be measured annually (i.e. independently of the
decenial censuses). One should not think it was an easy task. 
For instance, in the United states, in spite of the fact that
the earliest data, namely those for Massachusetts, go back
to the 19th century, it is only around 1930 that the so-called
birth and death ``Registration Areas'' were able to cover 
all the states. Basic as they are, these rates nevertheless
continue to raise unsolved questions. Here are two examples.\qL
Regarding birth rates, successive prediction attempts 
put foward in
past decades regularly proved wrong by a wide 
margin (of 15\% or more).
\qpar

For death rates the situation is different because the causes
of death are well known which means that, in contrast
to birth rates,
there are no unexplained fluctuations. What remains mysterious 
however is the overall pattern of death rates as a function of age.
Over the age of 15 years, the death rate follows the
well defined empirical law discovered by Benjamin Gompertz in the
19th century (Gompertz 1825). 
This law states that the death rate is an 
exponential function of age; basically it doubles when age
increases by 10 years. For ages under 15 there is also a 
well-defined law; it states that the death rate $ \mu $
{\it decreases} with age like an hyperbola function:
$ \mu \sim 1/t^{\alpha} $, where $ t $ is the age and $ \alpha $
an exponent which is close to 1. In previous studies
(Berrut et al. 2016, Bois et al. 2019) 
we have shown that this law also holds for non-human
multicellular organisms. It is of great interest to see
if it holds also for unicellular organisms for the following
reason. In multicellular organisms one can distinguish 
separate organs, with death usually occurring when one or several
crucial organ fail. In bacteria the situation is less
clear. Thus, this test may give us information
on the internal organization of unicellular organisms.
However, before turning to age-specific
death rates
we wish to give fairly accurate estimates of the
global death rate. 

\qA{Outline of the paper}

The paper proceeds as follows.
\qbu In the next section we describe the methods and results
obtained in previous studies.
\qbu In section 3 we carefully discuss and define the
death rate of a population of bacteria.
\qbu In section 4 we explain the protocol of our measurements,
we present our results and we discuss how their accuracy
can be improved.
\qbu In the last section we discuss related perspectives.

\qI{Previous studies of death rates of bacteria in exponential growth}

The investigation of the death rate of bacteria in their phase of
exponential growth has been going on for over a century. 
This long-lasting sustained interest underlines 
that the question is of
importance. At the same time the fact that it is not yet
completely solved suggests that this measurement
still represents an experimental
challenge. That is mostly due to the fact that the death rate is 
small. Novadays we know that for {\it E. coli} 
the death rate is of the
order of 1 per 1,000 living cells and per hour. \qL
In order to identify the trend of the death rate across the
exponential phase and in the early stationary phase requires
an even higher accuracy because the error bars must be smaller
than the effect of the trend itself.
\qpar

Actually the very definition of the death rate is not obvious.
Even if one leaves apart the question of the biological
meaning of the notion of ``death'', it is not obvious
to adapt standard demographic notions to
a population which doubles every 20mn. 
\qpar

In the following subsections it will be seen that there
are basically three methods for distinguishing
dead from living cells. 
\qbu The CFU (Colony Forming Unit) method.  The fact
that this traditional
method does not have the required accuracy is clearly shown
in Wilson (1922), a paper which will be discussed in more
details below.
\qbu The second method consists in following all individual cells
belonging to one cohort over as many  generations as possible.
Introduced in 1932, this method was vastly improved in
a paper of 2005 by Stewart et al. 
\qbu Finally, the third method is based on Flow Cytometry (FC).
It is this method that will be used in the present paper.

\qA{The Wilson paper (1922)}

In his paper Wilson
called into question the dogma that in a young broth
culture (up to 24 hours)
all bacteria are living. Here is what he writes at the beginning
of his paper.
\qdec{``On looking up the literature it was found that of 
the many observers [the author cites 9 papers published
between 1898 and 1920]
who had made a comparison of the two counts
[namely the total number of cells 
that can be counted in a culture, whether dead or alive
on the one hand
and the number of viable cells defined as cells that 
are able to fission on the other hand], 
the discrepancy was passed over with
little comment.
Any discrepancy was attributed to ``errors.''}
\qpar

In a 41-page long paper the author 
examines one by one
all successive operations required in counting procedures.
He tries to make them as rigorous as possible and he estimates
the remaining error margin. 
For instance dilution before counting is an operation
which requires great care but nevertheless cannot be made very
precise. In fact, the author was facing an impossible task. 
Such global counting techniques were beset
by too many uncertainties.
Therefore, it is not surprising that Wilson arrives at vague and
disappointing conclusions (p.444). 
\qdec{``It seems that in cultures of {\it Bact. suipestifer}
there is a normal death rate even during the period of maximum
growth. Its extent will vary from culture to culture.
In some it is as high as 43\%, in others it is only 20\% or 10\%,
while finally in a few it is for a short period actually nil.}
\qpar

\qA{Paper of Kelly and Rahn (1932)}

The first lines of the paper 
provide a clear statement of why
this is a key-issue. 
\qdec{``It has been assumed by many bacteriologists that
during the period of rapid growth, in a satisfactory 
culture medium, some bacteria {\it will die} in spite of
good food and favorable environment. No doubt this assumption 
was derived from an analogy with populations of higher
forms of life, of which a number of individuals are
known to die before they reach the reproductive age even with
good care.'' (Kelly et al. 1932, p.147)}
\qpar

The observation of
individual cells pioneered by Kelly and Rahn  
represented
a breakthrough. It is true that they
were unable to see any death in
the exponential phase but that was only because their
sample of 1,766 cells
was too small. Their methodology was sound and
opened the door to further observations 
with larger samples. Yet, this
did not happen until 73 years later. Indeed, in 2005 this
investigation was resumed.
Thanks to modern computerized counting techniques
and a sample some 20 times larger 
namely 35,000 {\it E. coli} cells, the experiment
revealed some 16 deaths; see Stewart et al. (2005)
and the discussion below.
.
\qpar

Using a method pioneered by J. Orskov (1922), Kelly and Rahn
followed the replication of individual bacteria 
(and one yeast species)
on a solid medium. They recorded the family trees spanning 
4 generations. Altogether they observed 1,766 divisions:
977 of {\it Bacterium aerogenes}, 325 of {\it Bacillus cereus}, 
464 {\it Saccharomyces ellipsoideus}.  On average the time intervals 
between fissions of  {\it Bacterium aerogenes} was 30 mn but with great
inter-individual fluctuations (coefficient of variation 
$ \sigma/m $ close to 100\%). 
\qpar

The observations led the authors to the following conclusions.
\qbu There was not a single instance where a cell 
became dormant.
\qbu In the very words used by the author, there was not
a single case of ``infant mortality'', that is to say 
a cell which died after division. 
\qpar

The first conclusion is reassuring because, even today, 
researchers often worry about the risk of mistaking
dormant cells for dead cells (see the discussion
at the begining of Garvey et al. 2007).
\qpar

The second conclusion is also interesting for it shows
that if such an infant mortality exists (as is indeed
observed in experiments done in the
past two decades as reported below)
then its rate is lower than: 1/1766 = 0.56 per 1,000.
(for a time interval of 30mn). This is a rough estimate
obtained by bulking together all three species. 
Estimates detailed by species (at least for the two largest)
are given in Table 1.
\qpar

Whereas the measurement of Stewart et al. (2005) relies
on the observation of individual cells, a paper of 2008
by Fontaine et al. relies on a global (not individual)
observation.
It is to the discussion of these modern investigations
that we turn now.

%
%% RECAPITULATIF DES 4 EXPERIENCES
\begin{table}
\centerline{Table 1: Landmark papers about bacterial death rates in the 
exponential phase}
\small
\vskip 2mm
\hrule
\vskip 1.5mm
\hrule
\vskip 2mm
$$\matrix{
 &\hbox{Year} & \hbox{Paper} \hfill  & \hbox{Method} \hfill & 
\hbox{Sample} & \hbox{Number} & \hbox{Death rate (dr)} \cr 
 & & \hbox{} & \hbox{} & 
 \hbox{size} & \hbox{of deaths} &\hbox{per 1,000} \cr
\qtb
 & & \hbox{} & \hbox{} & 
 \hbox{} & \hbox{} &\hbox{and per hour} \cr
\noalign{\hrule}
\qth
1& 1922 & \hbox{Wilson}  \hfill& \hbox{Global}\hfill & 
\hbox{several sets}  & \hbox{unreliable} &\hbox{unreliable} \cr
2& 1932 & \hbox{Kelly (1)}  \hfill& \hbox{Individual} & 
  \hfill 733 & <1 & \hbox{dr}<4.1 \cr
3& 1932 & \hbox{Kelly (2)} & \hbox{Individual} & 
  \hfill 420 & <1 & \hbox{dr}<1.4 \cr
4& 2005 & \hbox{Stewart} \hfill& \hbox{Individual} & 
 \hfill  35,049 & \hfill 16 & 1.5 \cr
\qtb
5& 2008 & \hbox{Fontaine} \hfill & \hbox{Global} \hfill& 
 \hfill 1,000,000  & \hfill 700 & 0.7 \cr
\noalign{\hrule}
} $$
Notes: ``Global'' means measurement performed on a 
large number of cells in liquid medium.
For this global measurement
it is the technique of flow cytometry which brought
about a breakthrough and allowed reliable measurements.
The following bacteria and yeast were investigated (in parenthesis
is the length in minutes of the reproduction cycle).
1: {\it Bact. suipestifer} and other species, 
2: {\it Bacterium aerogenes} (30mn),
3: {\it Saccharomyces ellipsoideus} (105mn), 
4: {\it E. coli} (30mn), 5: {\it E. coli} (30mn). 
\qL
Sources: Based on the papers cited in the third column.
\vskip 2mm
\hrule
\vskip 1.5mm
\hrule
\vskip 2mm
\end{table}
%%%%%%%%%%%%%%%%%%%%%%%%%%%%%

\qA{The paper by Stewart et al. (2005)}

In the Kelly and Rahn (1932) experiment
successive divisions were followed over
4 generations, a process which from each single initial cell
produced $ 2^4=16 $ cells.
In the Stewart et al (2005) experiment
up to 9 generations were followed, a process through which each
initial cell gave rise to $ 2^9 =512 $ {\it E. coli} cells.
Time-lapse images were taken and analyzed automatically thanks
to dedicated software. As 94 colonies were
analyzed this led to a total of 35,049 divisions. 
\qpar

The criterion used for the definition of death was immobility
combined with no division. Some 16 cell death were observed.
Unfortunately their sizes were not included in the publication
because the main purpose of the paper was a different issue.

\qA{Paper of 2008 by Fontaine et al. (FC measurement)}

In the technique that is used here
the cells are not monitored individually.
Instead, the recourse to flow cytometry allows global estimates.
Stained dead cells, 
or more precisely those cells whose breached membranes
allow the stain to drift into the cytoplasm,
are counted thanks to a flow cytometer (FC).
In such a device the light of a laser is diffused by the
cells when they move through the beam, then received by a sensor
and amplified by a photomultiplicator and finally analyzed by
a computer software algorithm.
Flow cytometry allows many characteristics of
the cells to be identified and
recorded. Here this technique is used to
count stained or fluorescent dead cells. 
\qpar

Flow cytometry
began to be used in the 1950s and was really a game changer.
It replaced the successive manipulations needed in the CFU method.
\qpar

Although FC is the key of the measurement method,
a number of additional verification tests are required
to ensure that what is measured by the device is indeed
the appropriate death rate.

\qI{Definition of the global death rate of bacteria}

In previously published mortality
data for bacteria (for instance in Steward et al. 2005) 
their
real meaning and definition, 
for instance with respect to the time interval, was
not clearly indicated. That is why we discuss this point
in some detail. 
\qpar

First, we recall the standard definition
of the death rate in human demography.
Then we examine how this definition should be adapted
to apply to organisms whose life duration is of
the order of 20 mn.

\qA{Global death rates in human populations}

In Appendix A we recall the reasons which lead to the 
definition of the death rate as accepted in human demography.
\qdec{Consider a population of size $ x $ that one observes
during a time interval $ \Delta t $. If $ \Delta y $ denotes
the number of deaths during $ \Delta t $, the death rate
in this time interval will be defined by:
$$ \mu(t)=1000\left( { 1 \over \Delta t }\right)
{ \Delta y \over x(t) } \qn{1} $$
where:
$ x(t) $ denotes the size of the population.
\qL
If $ \Delta t $ is expressed in a time unit $ u $, the unit
of $ \mu(t) $ will be  ``per 1,000 individuals and per $ u $''.}
\qpar

For a human population $ x(t) $ can be measured at the beginning,
mid-point or end of the one-year observation time because 
it changes little over one year. 
The situation is very different when one is dealing with 
a population whose size doubles every 20mn. In the next subsection
it will be seen that cell numbers should rather be replaced 
by life cycles.
\qpar

As an example of how to use the previous definition
let us assume a population of one million in which 120
deaths are counted in the month of January. If we take
years as unit of time, the interval of observation will be:
$ \Delta t= $ 1/12; thus, the January death rate will be 
$ \mu=1000\times 12\times 120\times 10^{-6}=1.44 $ per
1,000 individuals and per year.

\qA{Two special requirements for a population of bacteria}

For bacteria 
formula (1) must be applied with special care. Why? 
\qpar

Strictly speaking, $ x(t) $ should not be the whole population
existing at time $ t $ but only the individuals who are
subject to the risk of death during the time interval $ \Delta t $.
For instance, if one wishes to compute the death rate
of the French population in year 2000 one should not take
for $ x(t) $ the whole population alive on 31 December 2000
for this population would include individuals born on
the last day of 2000 and who are subject to the risk of dying
for only one day. Clearly, the number of such individuals
is negligeable compared to the whole population and one can 
therefore forget this restriction. However, with a population 
which doubles every 20mn this restriction must be taken 
into account. We will see below how to do that.
\qpar

There is another meaningful difference. To explain it, let us
consider a fictitious human population in which all individuals
live 100 years. We assume further that the initial
population is 1 million and that it
remains constant with births strictly compensating the deaths. 
Now, suppose we wish to estimate the death rate over a period
$ \Delta t $ of 300 years. This is basically the situation
we are facing with a sample of {\it E. coli}, except that 
here we assumed the 
population to be constant for the sake of simplicity.
In applying formula (1) should we take $ x(t) $ equal to
1 million? That would clearly be incorrect, for during
the 300 years there were 3 generations
successively exposed
to the risk of dying, each producing a number of dead
individuals which will be counted cumulatively.
In other words, instead of taking for $ x(t) $
1 million
one should take the number of life cycles
which is 3 millions. 
Therefore, when applying formula (1)
to {\it E. coli} one must also take for $ x(t) $ the number
of life cycles which occurred during the interval of
observation $ \Delta t $. 
\qpar

The number of life cycles for a population growing
exponentially is computed in Appendix A.

\qA{Global death rate of bacteria: modeling method}

The previous 
method required several assumptions to be made.
Can one devise a more direct argument? \qL
The present modeling method relies on two simple
equations which describe a population in exponential growth.
Any population can be fully described by 
two parameters: (i) the birth rate  and 
(ii) the death rate $ \mu $. In order to estimate them
one needs to connect them to two variables 
that can be measured in our experiment.
\qpar

For the birth rate we can start from the fact that the
population grows exponentially: $ x(t)=x_0\exp(\alpha t) $.
The exponent $ \alpha $ can be readily obtained from
the data of the spectrophotometer. It is true that this
device makes no difference between the cells which are
alive and those that are dead, but the second are
about 1,000 times less numerous than the first and can
therefore be neglected. 
\qpar

The cytometer provides the ratio $ y(t)/x(t) $ where
$ x(t) $ represents the number of
living cells and $ y(t) $ the number of dead cells.
We have already written an evolution equation for $ x(t) $,
can we also write one for $ y(t) $?
The evolution equation results from the definition
(1) written in differential form:
$$ \mu={ 1\over dt}{ dy \over x(t) } \rightarrow 
{ dy \over dt }=\mu x(t)=\mu x_0\exp(\alpha t) $$

If one assumes that $ \alpha $ is a constant, this equation
can be solved easily and one gets a relationship
between $ x(t) $ and $ y(t) $:
$$ y(t)-y_0={\mu x_0 \over \alpha }\exp(\alpha t)+K 
\quad \rightarrow \quad
y(t)-y_0={ \mu \over \alpha }[x(t)-x_0] $$
where $ x_0, y_0 $ are the initial values at the start of the
45mn time intervals between successive measurements.
\qpar

This equation implies that if $ x(t) $ growth exponentially,
then, so does $ y(t) $; therefore, the initial values 
can be neglected
in comparison to the values taken by $ x(t),y(t) $ some 45mn later.
$ x_0 \ll x(t=45 $ mn)  and $ y_0 \ll y(t=45 $ mn). 
Thus, the relationship between
$ x(t) $ and $ y(t) $ can be written in the simpler form:
 $$ \mu={ y(t) \over x(t) }\alpha \quad \hbox{where } 
\alpha = \log[x(t)/x_0]/t \qn{3}  $$

In practice, the ratio $ x(t)/x_0 $ can be read on the graph
of the optical density. Just to see what is the order
of magnitude of the death rate $ \mu $, we take an $ \alpha $
corresponding to a doubling time of 25mn which is the average in
our experiment. 
$ \alpha=\log(2)/0.42=1.66\ \hbox{h}^{-1} $ \qL
This gives:
 $$ \mu=1000\times 1.66{ y(t) \over x(t) } $$
\qpar

Five measurements of the ratio $ y(t)/x(t) $ 
covering the first 3 hours of the exponential phase, gave the
following average: $ <y(t)/x(t)>=0.0011\pm 0.00022 $ 
which in turn
leads to the following expression of the death rate: 
$$ \mu=1.83\pm 0.20 \hbox{ per 1,000 individuals and per hour} $$

%%%%%%%% EXPERIENCES

\qI{Methodology of the experiment}

The following description includes two parts that are often omitted
but which 
are of great importance for the success of our
experiment. The first is the selection of the micro-organism 
best suited
for our investigation. The second is the list of preliminary tests
performed to check for possible sources of errors in the successive
steps of the measurement process.  By listing
these sources of fluctuations our hope is also that it may help
to improve the accuracy of the observation.

\qA{Design of the experiment: selection of an appropriate strain}

As our objective is well defined we need to select the micro-organisms
which are best suited.
There are several requirements that we discuss below and which are
summarized in Table 2.
\qee{1}
The main requirement is a short division time so that one could
get a large population in a fairly short time.

%% Table: LIST OF CANDIDATES
%%-----------------------------------------------
\begin{table}[htb]

\small

\centerline{\bf Table 2\quad Characteristics of unicellular
organisms suitable for mortality studies}

\vskip 5mm
\hrule
\vskip 0.7mm
\hrule
%\vskip 2mm\hbox{}& \hbox{}\hfill & \hbox{type}\hfill & 
%\hbox{(hour)} \hfill & \hbox{formation} & \hbox{Eukariot}\cr

$$ \matrix{
\hbox{Name}& \hbox{Shape}\hfill & \hbox{Fission}\hfill & 
\hbox{Division} \hfill & \hbox{Size}&\hbox{Chain} & \hbox{Prokaryot}\cr
\hbox{}& \hbox{}\hfill & \hbox{type}\hfill & 
\hbox{time} \hfill & &\hbox{formation} & \hbox{or}\cr
\qtb
\hbox{}& \hbox{}\hfill & \hbox{}\hfill & 
\hbox{(mn, hour)} \hfill & \hbox{(micro-m)}&\hbox{} & \hbox{Eukaryot}\cr
\noalign{\hrule}
\qth
E.\ coli\hfill& \hbox{rod}\hfill & \hbox{transversal}\hfill & 
\hbox{20mn} \hfill & 1 &\hbox{little} & \hbox{P}\cr
Bacillus\ subtilis \hfill& \hbox{rod}\hfill & \hbox{transversal}\hfill & 
\hbox{20mn} \hfill & 1 &\hbox{yes} & \hbox{P}\cr
Saccharomices\ cerevisae \hfill& \hbox{round}\hfill & \hbox{budding}\hfill & 
\hbox{2h} \hfill & 3 &\hbox{yes} & \hbox{E}\cr
Euglena\ gracilis \hfill& \hbox{elastic}\hfill & \hbox{longitudinal}\hfill & 
\hbox{2h} \hfill & 20-100 &\hbox{yes} & \hbox{E}\cr
\qtb
Paramecium\ caudatum  \hfill& \hbox{long}\hfill & \hbox{transversal}\hfill & 
\hbox{20h} \hfill & 100 &\hbox{no} & \hbox{E}\cr
\noalign{\hrule}
} $$
\vskip 1.5mm
%Notes: 
%\qL
%Sources: \qL
\vskip 2mm
\hrule
\vskip 0.7mm
\hrule
\end{table}
%%--------------------------------------------------

% 
\qee{2}
After measuring the overall death rate
over a life cycle, in the second part of this investigation,
we are planning to measure the age-specific
death rate which means that we need a way to estimate
the individual age of each cell.
For rod-like
organisms experiencing transversal fission the bacteria 
length provides
a possible proxy for its age. The characteristics of diverse
model organisms summarized in Table 1 show that 
{\it Escherichia coli, Bacillus subtilis} and {\it Paramecium
caudatum} are three possible candidates. However, if one
wishes to count the dead cells with a cytometer, the last two
should be discarded because they may clog up the thin pipes
of the cytometer (Paramecium should anyway be discarded due
to its long generation time). This leaves us with {\it E. coli}.
\qee{3}
Another consequence of pipe narrowness
is the fact that one cannot use highly concentrated
cultures. In addition such cultures would not be suitable
because at the end of their exponential phase.
Therefore, if one wishes to test millions
of bacteria (in order to get several thousands of deaths)
one must use a cytometer which accepts
a large sample volume.

\qA{Organization of the experiment}

The successive steps of the experiment follow quite 
naturally from the end objective. They are summarized in Fig.1.
Every 45mn, a sample was extracted from the culture
which was growing in the incubator. Then, (i) its OD was measured
(ii)  the sample was diluted with PBS so as to have in each trial
the same cell concentration (iii) Propidium iodide was added
(iv) 1ml of the sample ws introduced in the cytometer. 
\qpar

In terms of time intervals, 
\qbu  It took about 10mn from sample
extraction to the PI step.
\qbu 15mn were given to the dead cells for absorbing
the PI stain. Note that the absorption is faster when 
the concentration of the stain is higher.  The PI
is said to be toxic but it would require a separate assay
to estimate the concentration level above which the
toxicity effect becomes notable.
\qbu The treatment of the 1ml sample in the FC took
about 10mn to 15mn. In fact, the value of the ratio dead/total
observed after 2 or 3mn remained very stable in the remaining
time. 
This means that a smaller volume would have given the same results.

%
%%%%% ORGANISATION DE L'EXPERIENCE
\begin{figure}[htb]
\centerline{\psfig{width=10cm,figure=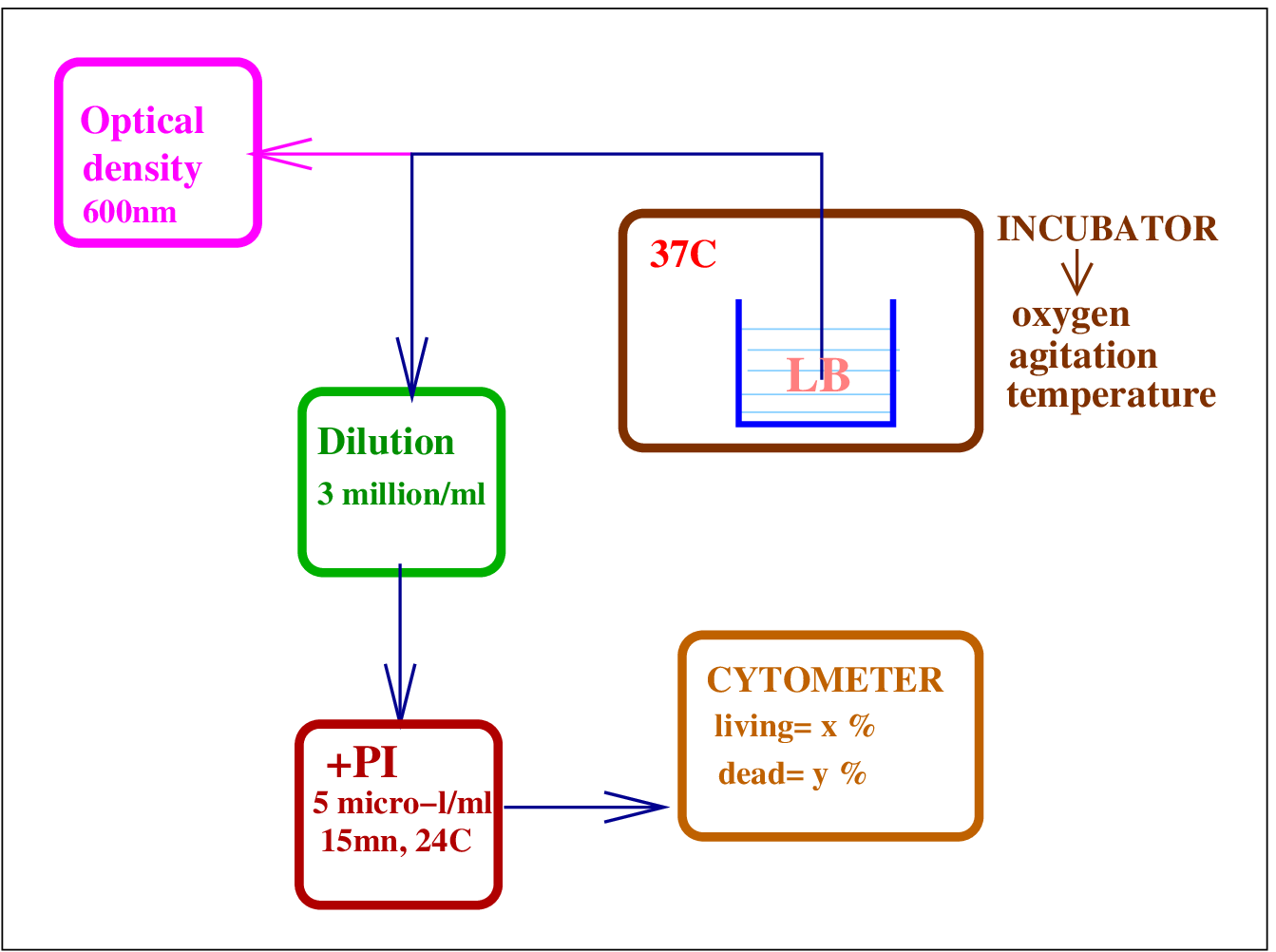}}
\qleg{Fig.1 Organization of the experiment.}
{The two main devices are the spectrophotometer which measures
the optical density (which in turn gives a measure of the number
of cells per cubic cm)  and the flow cytometer which counts
the number of dead cells after they have been made 
fluorescent by absorption of propidium iodide.
Note that the time interval between extraction of a sample from
the culture and its introduction in the cytometer is about 20mn.}
{}
\end{figure}
%-------------------------------------------------

\qA{Preliminary tests}

Our tests concerned mainly three points.
\qbu When does the stationay phase really begin? The transition
becomes clearly visible toward an optical density of 0.3
(for a more elaborate discussion see Sezonof et al. 2007). 
However, by starting a culture with
a concentration of cells some 1,000 smaller than the smallest
OD that the spectrophotometer could measure, we have seen that
there is in fact a 
steady increase of the doubling time.
\qbu In order to test the accuracy of the identification
of the dead cell we prepared a mixture 50\%--50\% of dead cells
(killed by a temperature shock) and living cells. \qL
A proportion of 45\% dead cells was detected by the FC.\qL
%{\color{red} chiffre a confirmer}
As a complementary test, the dead cells sorted by the FC were reintroduced
into the FC. A ratio dead/total of 97\% was observed. 
%{\color{red} chiffre a confirmer}
%
\qbu In an attempt to better understand the connection between
``PI death'' and ``immobility death'', dead cells sorted by the FC
were put under a microscope equipped with a light source in the
fluorescence range of the PI. It was verified that there remained
no moving cells in the sample. This test tells us that our ``PI dead''
would also be counted as dead with 
the immobility criterion used in the experiment
of Stewart et al. (2005). In other words, PI death is either
equivalent to or ``deeper'' than ``immobility death%
\qfoot{During an observation of the embryonic growth of eggs of
Zebra fish,
one embryo remained completely motionless during two days. It
was thought to be dead but in fact resumed a normal development.
Compared with normal embryos, the only difference was that this one
hatched two days later.}%
''.

\qI{Results}

\qA{Absolute versus relative measurements}

It must be stated from the outset that we are more interested
in relative than in absolute values of the death rate. There
are two main reasons for that.
\qbu The absolute value of the death rate is dependent upon
the criterion used to define ``death''. We have already mentioned
that PI death seems to be a sharper criterion than death defined
by immobility. In addition, even within PI death the technical
definition of death relies on how one selects the lower
limit of PI positive events. On Fig.2 this limit is represented
by the left-hand side of the rectangle which defines the 
PI positive events%
\qfoot{Also to be mentioned is the fact that the number of living
cells is affected  by the background noise of the cytometer.
This is seen very clearly through a test-run in which only
PBS+LB (without cells) is introduced in the FC. Then the
background noise appears in the form of a set of events at the
far left of the domain delimiting the living. This means
that the total number of living cells detected by the FC
is an overestimate. 
The purpose of setting an appropriate left-hand side limit for the
domain delimiting the living cells is precisely to eliminate these
spurious events.}%
.
Moreover, one should realize that the absolute level 
of the death rate  will not really
affect our understanding of the cells' mechanisms. 
\qbu On the
contrary relative death rates are of great significance. The
word ``relative'' can be given different meanings. \qL
In the present
paper our main interest is to see how the death rate behaves
when the culture moves from the exponential phase to the
stationary phase. Does it increase or does it fall%
\qfoot{Arguments in favor of a predicted fall were given in
a recent separate paper (Di et al. 2021).}%
?
In the framework of our investigation a second possible meaning 
is to determine whether there is an ``infant
mortality effect'' in the terms of Kelly and Rahn's paper.
It means comparing the death rates of 
newly divided versus mature cells. We expect a death rate peak for
freshly divided cells followed by a steady decline as the
size of the cells increases. 
%
%%%%% EVENEMENTS IP POSITIFS
\begin{figure}[htb]
\centerline{\psfig{width=12cm,figure=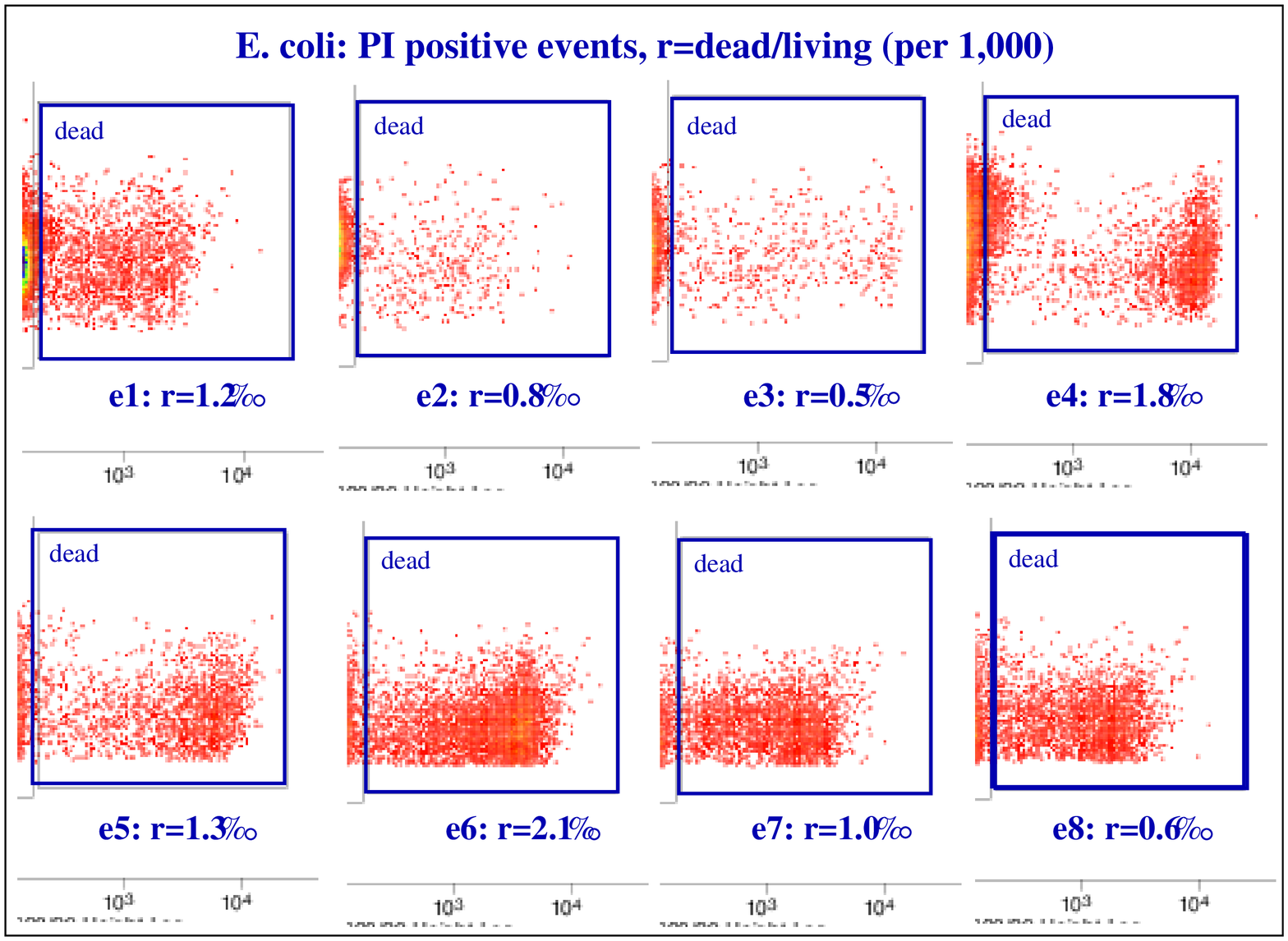}}
\qleg{Fig.2 PI positive events for 8 samples
extracted at 45mn time intervals.}
{The scale of the x-axis gives the intensity of PI fluorescence
as detected by the cytometer. 
In spite of the fact that the same procedure was implemented
in the successives assays, the distributions show 
notable fluctuations.
Needless to say, these fluctuations affect
the estimates of the number of dead cells. It will be the purpose
of forthcoming experiments to determine whether these fluctuations can
be reduced or whether they are purely random fluctuations.}
{}
\end{figure}
%-------------------------------------------------

\qA{Graph of dead/living}
%
%%%%% GRAPHE DE LA DO ET DE y/x
\begin{figure}[htb]
\centerline{\psfig{width=14cm,figure=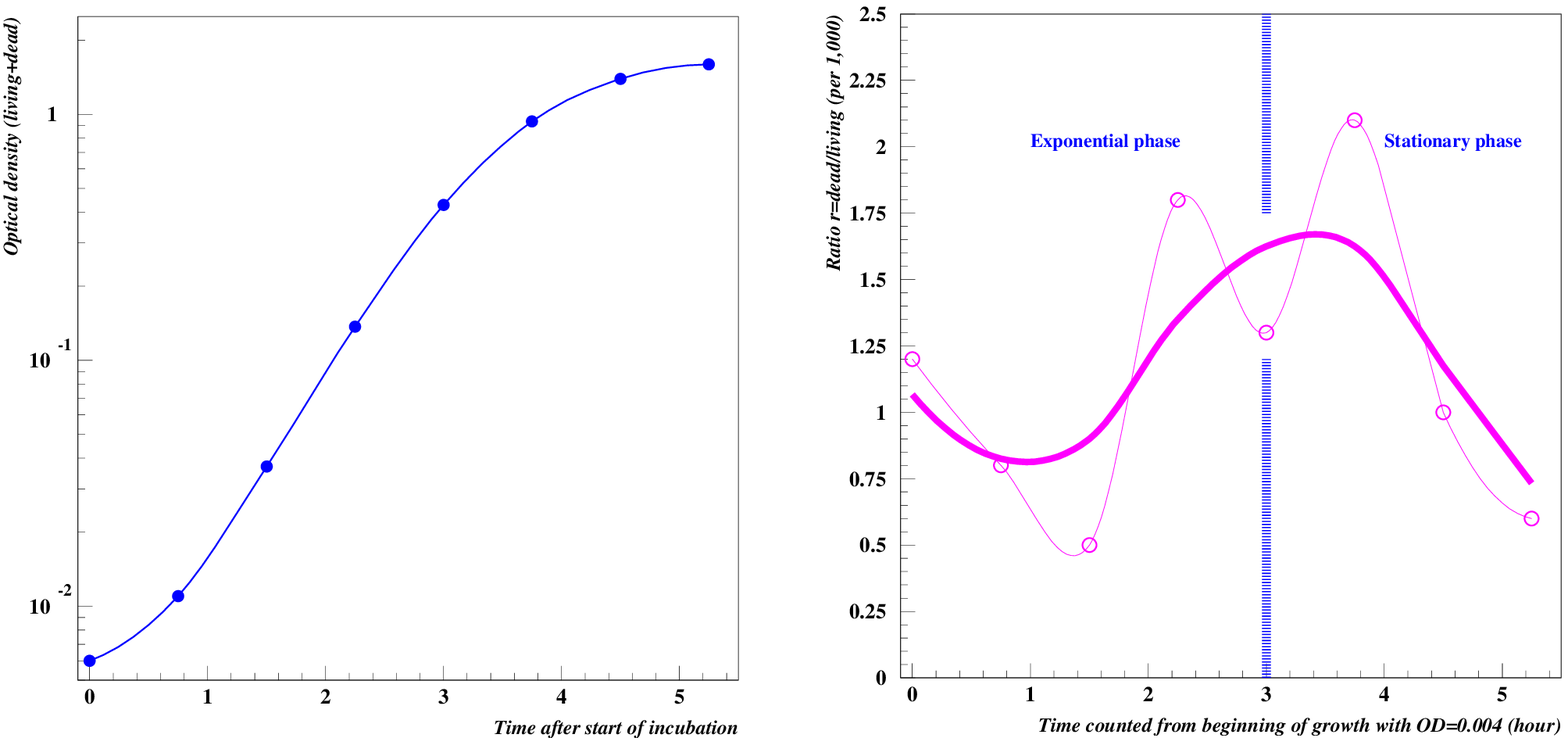}}
\qleg{Fig.3a,b Optical density and ratio dead/living for 8 samples
taken from the culture at 45mn time intervals.}
{(a) This growth curve serves to define the two regions
labelled as ``Exponential phase'' and ``Stationary phase'' in
Fig.3b.\qL  
(b) This curve gives the ratio dead/living which is 
denoted $ r=y/x $ in the text. From this ratio it is easy
to derive the death rates by using formula (4) given
in the text. Note that whereas $ r $ is a number without
dimension, for the death rate one must specify the time unit.}
{}
\end{figure}
%-------------------------------------------------

We chose to represent the ratio $ y/x $ rather than the 
death rate itself for two reasons.
\qbu Whereas the ratio $ y/x $ is read directly on the cytometer,
the computation of the death rate relies on a set of assumptions
described in the text. It is true that
the overall level of $ y/x $ depends
upon parameter selection and fine tuning of the cytometer, but the 
{\it relative variations} of $ y/x $ across the 8 measurements performed
over a period of five and a half hours are independent of
how the parameters of the cytometer have been selected.
\qbu Whereas
the absolute level of the death rate may be useful for
the purpose of inter-species comparisons, one should realize
that anyway
such comparisons are necessarily hazardous due to the
fact that even for close species such as {\it E. coli} and
{\it B. subtilis} there are many ingrained differences.
\qpar

Is there a fall in the death rate $ \mu(t) $ in the course of time
as was predicted in Di et al (2021)? The fluctuations are
somewhat too large to permit a clear conclusion. However, it
appears  that there is no major increase. 
\qpar

{\bf Remark} If the ratio $ y/x $ is considered
to be approximately steady, formula (3) shows that
$ \mu(t) $ is proportional to the growth exponent $ \alpha $ of the 
population. As $ \alpha $ decreases with $ t $ it might seem
that $ \mu $ will also fall with $ t $. Unfortunately, this
reasoning is not correct for in solving the differential equation
$ \alpha $ was supposed constant. 
\qpar

{\bf A note about fluctuations} Fig.3b might suggest a fairly
high level of fluctuations. In reality, the absolute level of the
fluctuations are of the order of one per 1,000 which
means that, compared to similar observations,
they are in fact of same order. Here is a comparison.
\qpar

Using a hemocytometer, Jones et al. (1985, p.78, table 1)
determined
the proportion of viable spleen cells of mice
through a staining method using
a mixture of propidium iodide and fluorescein diacetate.
In order to assess the accuracy of their method they repeated the
measurements 8 times. The coefficient of variation (standard deviation
divided by the mean) of the 8 proportions  was 40\%.
For the 8 measurements represented
in Fig.3b the coefficient of variation is 51\%. In other words,
it is only because the proportion that we measure is quite small
that the fluctuations depicted in Fig.3b appear to be large. 
\qpar

In addition the graph shows 
that in order to clearly identify a 
downward trend, the fluctuations would have to be further reduced
by a factor 2 or 3.

\qI{Conclusion and perspectives}

While still open to improvement, our death rate measurements 
appear more accurate than those of Stewart et al. (2005) and
Fontaine et al. (2008). To be fair one should add that the 
death rate measurement was not the main
goal of these papers. \qL
The important point in Fig.3b is not so much the absolute
level but rather the magnitude of the fluctuations. Although
already faily small, the fluctuations will have to be reduced
further in order to see if a trend can be identified. In addition
it would be appropriate to add two or three data points in the
stationary phase that is to say above the optical density of 1. 
\qpar

Although the cytometric measurements must be handled with great care
they 
appear to be the only promising method for this kind of 
bio-demographic investigations.

\appendix

\qI{Appendix A. Death rates in human and bacterial populations}

\qA{Global death rates in human populations}

For a human population the global death rate of a 
population $ A $ is obtained by first counting the number
of deaths
in one year (that we denote $ \Delta y $) and by dividing this
number by the size, $ x(t) $, of the population.
The ratio $ r_d=\Delta y/x(t) $ represents
the relative fall of the population.
\qpar

However,  this definition 
does not give a satisfactory death rate. The reason can be seen
by considering neonatal mortality data. Usually they are
given for the following successive intervals: (i) less than
one day, (ii) from one day to less than one week (iii) from 8 days
to less than 4 weeks. Needless to say, $ \Delta y $
will be heavily dependent upon the respective time intervals:
for (i) it is one day, whereas for (iii) it is 28-7=21 weeks. 
With a stable number of deaths on successive post natal days
the $ r_d $
of interval (iii) would artificially appear 21 times larger
than for interval (i). If the number of deaths on successive days
are not the same (as is indeed the case) this phenomenon
will be completely hidden by the change in time intervals.
This difficulty can be overcome easily by normalizing $ r_d $
by the length of the time interval, thus instead of
$ r_d $ one should consider: $ \mu=r_d/\Delta t $.
With this definition identical deaths in successive days
will indeed lead to a constant $ \mu $. This $ \mu $
can be taken as an appropriate definition of the
death rate. \qL
The present normalization argument could appear somewhat trivial, 
but actually
in many national data for neonatal death rate statistics,
this normalization is {\it not} made.
\qpar

Because, $ x(t) $ is usually much larger than $ \Delta y $,
it is often convenient to multiply $ \mu $ by 1,000 or
equivalently to express $ x(t) $ in thousands of individuals.
\qpar

In addition to the time length of an observation $ \Delta t $
just introduced, one must also specify which time unit ($ u $)
is used. This did not play any role in the previous argument
because implicitly it was the day which was selected as time unit.
In order to see the role of $ u $ consider the following case.\qL
One wishes to measure the death rate between two censuses.
In this case, $ \Delta t $ is 10 years; if for $ u $
we take also 10 years, $ \mu $ will give
the death rate per 1,000 and per
10 years. Usually, this is not what one wants because one wishes
to compare $ \mu $ to the annual death rate. To obtain
an annual death rate, one should take $ u= $ one year. This will
give a death rate that is one tenth of the previous one. It 
represents the average annual death rate over the 10-year period. 
\qpar

\qA{Estimates of the number of life cycles}

In our experiment the number of dead {\it E. coli}
cells were measured every 45 mn. 
However for the sake of simplicity of the present
argument we assume that the measurements were
performed every 60 mn. Assuming in addition a
generation time of 20 mn, there will be 3 doublings.
If 
we assume that they were 10,000 
bacteria at initial time $ t_0=0 $, 
at the end of the 60mn there will be 80,000.
\qpar

At time $ t_1=40 $mn the population
will number $ 40,000 $. All these 40,000 cells 
were born before
$ t_1 $ and all of them had been
exposed to the risk of death during their whole
life cycle. 
\qpar

If we assume that all cells are synchronized they will all
divide at time $ t_1 $ and the resulting 80,000 cells will be
exposed to the risk of dying during their whole life cycle.
In this case the number of life cycles would be:
$$ N_l=20,000+40,000+80,000=80,000(1/4+1/2+1)=1.75x(t) $$

where $ x(t) $ is the end population.
\qpar

However, in reality the cells are not synchronized. Hence the previous
formula should be changed according to the following discussion.
It turns out that the formula $ N_l=1.75x(t) $ should be replaced
by: $ N_l=1.37x(t) $. 
\qpar

Cells 
born at $ t_2=50 $mn, will be exposed to the risk of death
for only 10mn, that is to say one half of the life cycle;
those born after $ t_2 $ will be exposed to the risk
for an even shorter time. Clearly, to continue this argument
one needs to know the distribution by age of the 
third generation.
\qpar

One can get an idea of
the distribution of {\it E. coli} cells by age at a 
given moment through
the distribution of their lengths. Examples of distributions
by size are given in the literature 
and were found
in agreement with our own measurements. They show a 
substantial peak in young age followed by a slow and fairly steady
fall. For the sake of simplicity this kind of distribution
will be schematized as having two separate peaks: one at $ t=45 $mn 
corresponding to 3/4 of the generation and a second at $ t=55 $mn
which corresponds to the remaining 1/4. Those of the first
peak will be exposed to the risk of death for 15mn, i.e. 0.75
life cycle, whereas those of the second peak will be exposed
for only 5mn, i.e. 0.25 life cycle. \qL
Thus, the total number of life cycles  subject to the risk 
of dying can be written:
$$ N_l = 20,000 +40,000+
80,000[(3/4)\times 0.75 + (1/4)\times 0.25] $$
which gives:
$$ N_l=80,000\times (0.25+0.5 + 0.62)=1.37x(t) $$

Now, if one assumes that $ n_d $ dead cells have 
been counted by the cytometer
we can apply definition (1) with:
$$ x(t)=N_l \quad \Delta y=n_d\quad \Delta t=1\hbox{hour}\quad
u=\hbox{hour}  $$

Replacement leads to the following death rate:
$$ \mu=1000{ 1\over lc }{ n_d \over 1.37x(t) }=
1000\times 0.82r  $$

where $ lc $ denotes the length of a life cycle and
$ r $ denotes the ratio dead/living given by the cytometer.
\qpar

Note that the cytometer will also count a number of
dead cells whose death occured before the 45mn-long observation time;
but due to the exponential growth this number can be 
considered negligeable
compared to the deaths which occurred during the subsequent interval
of observation.
\qpar

If in the previous argument we replace the time interval of
60mn by the one of 45mn which was actually used in our experiment
the same argument leads to the following expression for
total number of life cycles:
$$ N_l=23,784+47,570[(3/4)\times 0.75 + (1/4)\times 0.25] =1.12x(t) $$

In writing this expression we have separated the time interval
of 45mn into one of 25mn and a second of 20mn. In 25mn the
population will reach $ 2^{25/20}=23,784 $ and all these cells
will be able to finish their live cycle. In contrast, the cells
born during the last 20mn will not be able to finish their life
cycle and will give rise to a number of life cycles that can be
estimated by the same argument as above. 
\qpar
This leads to the following expressions for the number of life
cycles and for the death rate: 
$$ N_l=1.12x(t)\ \longrightarrow \ 
\mu={ 1\over lc }{ n_d\over 1.12x }=1000\times 1.19r  \qn{2}  $$

\vskip 6mm

{\bf References}

\qparr
Berrut (S.), Pouillard (V.), Richmond (P.), Roehner (B.M.) 2016:
Deciphering infant mortality.
Physica A 463,400-426.

\qparr
Bois (A.), Garcia-Roger (E.M.), Hong (E.), Hutzler (S.),
Irannezhad (A.), Mannioui (A.), Richmond (P.),
Roehner (B.M.), Tronche (S.) 2019:
Infant mortality across species. A global probe of congenital
abnormalities.
Physica A 535,122308.

\qparr
Di (Z.), Garcia-Roger (E.), Richmond (P.), Roehner (B.M), 
Tronche (S.) 2021: Transient frailty induced by cell division.
Reasons and implications.
Preprint available on arxiv.

\qparr
Fontaine (F.), Stewart (E.J), Lindner (A.B.), Taddei (T.) 2008:
Mutations in two global regulators lower
individual mortality in {\it Escherichia coli}.
Molecular Microbiology  67,1,2-14.

\qparr
Garvey (M.), Moriceau (B.), Passow (U.) 2007: Applicability of
the FDA stain [fluorescein diacetate] assay to determine the
viability of marine phytoplankton under different environmental
conditions. 
Marine Ecology Progress,  series 352,17-26\qL
[The paper contains impressive pictures of stained cells.]

\qparr
Jones (K.H.), Senft (J.A.) 1985: An improved method to determine
cell viability by simultaneous staining with fluorescein 
diacetate -- propidium iodide.
The Journal of Histochemistry and Cytochemistry 33,1,77-79.

\qparr
Kelly (C.D.), Rahn (O.) 1932: The growth rate of individual
bacterial cells.
Journal of Bacteriology 23,3,147-153.

\qparr
Orskow (J.) 1922: Method for the isolation of
bacteria in pure cultures from single cells and procedure for the
direct tracing of bacterial growth on a solid medium.
Journal of Bacteriology 7,537-549.

\qparr
Sezonof (G.), Joseleau-Petit (D.), D'Ari (R.) 2007: 
{\it Eschrichia coli} physiology in Luria-Bertani broth.
Journal of Bacteriology December 2007,8746-8749.\qL
[The authors propose a detailed investigation of the
exponential growth phase of {\it E. coli} which, following the
lag phase, commonly lasts of the order of two hours.]

\qparr
Stewart (E.J.),  Madden (R.), Paul (G.), Taddei (F.) 2005:
Aging and death in an organism that reproduces by
morphologically symmetric division.
Plos Biology 3,2,e45.

\qparr
Wilson (G.S.) 1922: The proportion of viable bacteria in
young cultures with especial reference to the technique
employed in counting.
Journal of Bacteriology 7,4,405-446.

%%%%%%%%%%%%%%%%%%%%%%%%%%%%%%%%%%%%%%%%%%
%%%%%%%%%%%%%%%%%%%%%%%%%%%%%%%%%%%%
%%%%%%%%%%%%%%%%%%%%%%%%%%%%%%%%%%%%
\count101=0  \ifnum\count101=1

%%% CE QUI SUIT FUT ECRIT APRES L'EXPERIENCE DE JUILLET.
%%% CE FUT LA PREMIERE OU L'ECHANTILLON FUT PREPAREE
%%% SELON LA PROCEDURE D'ARNAUD. APPAREMMENT, CETTE
%%% PRECEDURE ETAIT MEILLEURE QUE LA MIENNE CAR TX DE
%%% DECES PLUS FAIBLE (AUTOUR DE 1/1000)

Why do we think that there is a mortality spike shortly
after division? 
\qpar

One reason is that this has been observed 
previously in all our studies of various multicellular organisms.
This cannot be taken as
a sufficient reason, however. Even fairly
primitive organisms (e.g. rotifers) have organs (although
rotifers have only some 1,000 cells, they have
digestive, neural and reproductive systems) whose failure
means death. Bacteria have also a number of distinct components
but we are less familiar with how their failure affects
the cell. We come back to this point later on.
\qpar

A second argument is that the replication process represents
the bulk of the manufacturing process which leads to 
the creation of a new individual through binary 
fission. In this process two parts are usually distinguished,
namely (i) {\it karyokinesis} which designates the DNA replication,
and (ii) {\it cytokinesis} which refers to cytoplasm division.

The present study is the continuation of several
foregoing investigations of mortality rates in the early
moments after the birth of a new organism. 
The expression ``infant mortality'' is used in the
field of reliability science to designate early failures
due to manufacturing defects. It has also been used
in a paper by Kelly and Rahn 
published in 1932 which was one of the first studies
which tried to estimate the mortality rate in the exponential phase
of bacterial growth. Because it was based on small samples
of less than 1,000 individuals this study did not find any
death. Nevertheless, it provided the useful conclusion,
that if it exists,
the mortality should be smaller than 1 per
1,000 and per hour.
Along with data found in more recent papers
going in the same direction (see below), this information was
instrumental in helping us to design our own experiment.
\qpar

How did we design the planned experiment? 
There are several requirements.
\qee{1}
The main requirement is a short division time so that one could
get a large population in a fairly short time.
\qee{2}
After measuring the overall death rate
over a life cycle, in the second part of this investigation,
we wish to measure the age-specific
death rate which means that we need a way to estimate
the individual age of each cell.
For rod-like
organisms experiencing transversal fission the bacteria 
length provides
a possible proxy for its age. The characteristics of diverse
model organisms summarized in Table 1 show that 
Escherichia coli, Bacillus subtilis and Paramecium
caudatum are three possible candidates. However, if one
wishes to count the dead cells with a cytometer, the last two
should be discarded because they may clog up the thin pipes
of the cytometer (Paramecium should anyway be discarded due
to its long generation time). This leaves us with E. coli.
\qee{3}
Another consequence of pipe narrowness
is the fact that one cannot use highly concentrated
cultures. In addition such cultures would not be suitable
because at the end of their exponential phase.
Therefore, if one wishes to test millions
of bacteria (in order to get several thousands of deaths)
one must use a cytometer which accepts
a large sample volume.

Another crucial requirement

%% Table: LIST OF CANDIDATES
%%-----------------------------------------------
\begin{table}[htb]

\small

\centerline{\bf Table 1 \quad Characteristics of unicellular
organisms suitable for mortality studies}

\vskip 5mm
\hrule
\vskip 0.7mm
\hrule
%\vskip 2mm\hbox{}& \hbox{}\hfill & \hbox{type}\hfill & 
%\hbox{(hour)} \hfill & \hbox{formation} & \hbox{Eukariot}\cr

$$ \matrix{
\hbox{Name}& \hbox{Shape}\hfill & \hbox{Fission}\hfill & 
\hbox{Division} \hfill & \hbox{Size}&\hbox{Chain} & \hbox{Prokariotic}\cr
\hbox{}& \hbox{}\hfill & \hbox{type}\hfill & 
\hbox{time} \hfill & &\hbox{formation} & \hbox{or}\cr
\qtb
\hbox{}& \hbox{}\hfill & \hbox{}\hfill & 
\hbox{(mn, hour)} \hfill & \hbox{(micrometer)}&\hbox{} & \hbox{Eukariotic}\cr
\noalign{\hrule}
\qth
\hbox{E. coli}& \hbox{rod}\hfill & \hbox{transversal}\hfill & 
\hbox{20mn} \hfill & 1 &\hbox{little} & \hbox{P}\cr
\hbox{Bacillus}& \hbox{sphere}\hfill & \hbox{diameter}\hfill & 
\hbox{20mn} \hfill & 1 &\hbox{yes} & \hbox{P}\cr
\hbox{subtilis}& \hbox{}\hfill & \hbox{}\hfill & 
\hbox{} \hfill & &\hbox{} & \hbox{}\cr
\hbox{Saccharomices}& \hbox{round}\hfill & \hbox{budding}\hfill & 
\hbox{2h} \hfill & 3 &\hbox{yes} & \hbox{E}\cr
\hbox{cerevisae}& \hbox{}\hfill & \hbox{}\hfill & 
\hbox{} \hfill & &\hbox{(hypha)} & \hbox{}\cr
\hbox{Euglena}& \hbox{}\hfill & \hbox{longitudinal}\hfill & 
\hbox{2h} \hfill & 20 \hbox{sphere} &\hbox{yes} & \hbox{E}\cr
\hbox{gracilis}& \hbox{}\hfill & \hbox{longitudinal}\hfill & 
\hbox{2h} \hfill & to 100 \hbox{flat} &\hbox{(in light)} & \hbox{E}\cr
\hbox{Paramecium}& \hbox{long}\hfill & \hbox{transversal}\hfill & 
\hbox{20h} \hfill & 100 &\hbox{no} & \hbox{E}\cr
\qtb
\hbox{caudatum}& \hbox{}\hfill & \hbox{}\hfill & 
\hbox{} \hfill &  &\hbox{} & \hbox{}\cr
\noalign{\hrule}
} $$
\vskip 1.5mm
Notes: 
\qL
Sources: \qL
\vskip 2mm
\hrule
\vskip 0.7mm
\hrule
\end{table}
%%--------------------------------------------------

\qI{Death rates in a multi-generation growth process}

\qA{Position of the problem}

In human demography death rates are usually computed
annually that is to say over time intervals much shorter
than a generation (i.e. some 20 years). On the contary, 
an experiment on E. coli may extend to as many as 7 or 8
generations.
As this is an uncommon situation it requires 
revisiting the standard framework. 
\qpar

More specifically, for a population of size $ x(t) $
at time $ t $ the rate of change 
$ \lambda $ is classically defined as:
$$ \lambda(t) =\left( { 1 \over x(t) }\right){ dx \over dt } \qn{1} $$

This is the differential definition of the rate which means that
$ dt $ and $ dx $ are supposed to be arbitrarily small.
Geometrically, $ \lambda $ is the slope of the curve 
$ \left(t, \log[x(t)] \right) $
at time $ t $. 
It defines a growing or a falling population depending
on whether $ dx $ is positive or negative.
Naturally, when this definition is used in a real case, $ dt $
and $ dx $ become finite quantities, which means that the 
definition (1) becomes:
$$ \lambda(t) =\left( { 1 \over x(t) }\right)
{ \Delta x \over \Delta t } \qn{1'} $$

At first sight the change from (1) to (1') may seem unimportant
but, as will be seen below, it creates serious difficulties.
The relation (1') shows that in order to compute $ \lambda $ one
needs to know three quantities: 
(i) the time $ \Delta t $ during which
the population is exposed to the change process, \qL
(ii) the change $ \Delta x $ of the population, \qL 
(iii) the population $ x(t) $.
\qpar

The troubles steem from this last factor: what should
we take for $ x(t) $? Should it be the population (i) at the
beginning of the interval $ I=(t,t+\Delta t) $ or (ii) at its
end, or is it ``better'' to take for $ x(t) $ (iii) the average 
$ \overline{x(t)} $?. 
When $ \Delta t $ is one year, the matter
is not very serious for a human population changes only
a few percent within one year. 
Actually, however, 10 years is a more natural interval because
in most countries censuses take place every 10 years which means
that the population is really measured only every 10 years,
the figures in between are estimates. We will see below that
depending on whether one uses (i), (ii) or (iii)
makes a substancial  difference of more than 10\%.
\qpar

Ten years represents only one half of the time corresponding 
to a human generation. In
our E. coli experiment we measure the deaths accumulated over
7 or 8 generations. In other words, the exact definition
of the death rate becomes here of paramount importance.
If we do not define precisely what
option we are using for $ x(t) $ and why,
then it will be impossible to compare our results to those
which will be obtained in our upcoming studies or 
in those obtained by other 
researchers.
\qpar

In the following subsections we examine two phenomena
that are crucial in our experiment, namely the 
rapid exponential growth of E. coli
and how to estimate the rate of the mortality surimposed 
on the exponential growth. 

\qA{Growth rate of a rapid exponential process}

Exponential growth is a simple case in the sense 
that one knows the shape of the growth function, namely:
$$ x(t)=x_0\exp(\lambda t) \rightarrow \log\left[x(t)/x_O\right] =
\lambda t \qn{2} $$
 
which means that in a semi-log representation the growth
function will be a straight line whose slope gives an estimate
of the growth rate $ \lambda $. 
Fig.1 shows
the growth functions observed in our experiment.

%
%%%%% LES CROISSANCES EXPONENTIELLES
\begin{figure}[htb]
\centerline{\psfig{width=16cm,figure=semilog.eps}}
\qleg{Fig.1 Exponential growth of three populations
starting from different concentrations of E. coli.}
{Note that, strictly speaking, 
the numbers given by the spectrophotometer
provide estimates of {\it all} cells, whether living or
dead. However, based on previous studies (see text)
one expects
the dead cells to be a small fraction (of the order of 1/1000) 
of the living cells.
The doubling times (and growth rates per hour
within parentheses)
of the three curves are,
from bottom to top: 20.6mn (2.04),
22.0mn (1.92), 
24.3mn (1.73)
respectively.
We see that the lower the initial E. coli density, the
faster the initial growth.}
{ }
\end{figure}
%-------------------------------------------------

Instead of $ \lambda $, the exponential growth
process can also be described by its {\it doubling time} $ \tau $.
The two parameters are related as follows:
$ \tau=\log 2/\lambda $.

\qA{Exponential growth of the US population, 1800-1900}

In order to illustrate the difficulty due to different 
possible options (as outlined above)
we consider the
population of the United States between 1800 and 1900.
The growth from
5.3 millions in 1800 to 76 millions in 1900 was approximately
exponential with a global growth rate
$ \lambda=(1/100)\log(76/5.3)=2.66\% $ per year, which in turn
leads to a doubling time of: $ \tau=\log 2/\ 2.66=26.0 $ years.\qL
For a growth that is strictly exponential, one would get the
same $ \lambda $ estimate when computing it on sub-intervals
In fact, the real values fluctuate around the global average.
For instance, between 
1830 and 1840 one gets: $ \lambda(t_1)=2.84\% $ whereas from
1870 to 1880 $ \lambda(t_2)=2.74\% $, a relative 
difference of 3.6\%.
\qpar

Let us now compare the estimates obtained through the three options
described above. As an example we take the decade 1830 to 1840
which witnessed a growth from 12.8 to 17.0 millions. 
One gets: 
$$ \lambda_{i}=3.28,\quad \lambda_{ii}=2.47,\quad \lambda_{iii}=2.82 $$

These rates show a spread of 33\%, i.e. ten times more than the 
time fluctuations of 3.6\% seen above. 

\qA{Counting the deaths}

So far we have only considered population growth, but
in any growth process there are also
some individuals who die. The deaths may be 
a small fraction of the living (as is indeed the case in our
experiment)
but if the population is large enough they will 
become visible and countable. 
\qpar

When all individuals of a growing population of E. Coli
are followed in the course of time, as done
in the studies of Kelly and Rahn (1932) and Stewart et al. (2005),
the deaths can be counted chronologically as they occur one by one.
This situation is similar to 
death counting in a human population.
\qpar

On the contrary in our experiment the deaths occurring
in a given time interval $ (0,t) $ (of the order of one hour
and a half) will be made visible only at time $ t $
through the addition of
propidium iodide so that the cytometer can count them.

In the next subsection we write the equations which summarize
this process.

\qA{The equations which rule the death numbers}

We assume here that, as is the case in our experiment, 
we are able to count
the cumulative number of deaths, denoted by $ y(t) $ that occur 
in the growing population of E. coli. The number of the living cells
will be denoted by $ x(t) $. As there  two unknown functions 
$ x(t),y(t) $, to describe this process we need two equations.
\qpar

The spectrophotometer allows us 
to estimate the total number (i.e. living plus dead)
of cells. From Fig.1 we know that it 
is an exponential function and the graph also gives an estimate
of the death rate $ \lambda $.
This gives us our first equation:
$$ x(t)+y(t)=c_0\exp(\lambda t) \hbox{ where } c_0=x_0+y_0 \qn{3} $$

In equation (3) $ x_0 $ and $ y_0 $ denote the ``initial'' densities
of living and dead cells; $ c_0 $ can be read on Fig.1 for $ t=0 $.

\appendix

\qA{Estimating the death rate in an exponentially growing population}

\fi

\end{document}